\documentclass[aps,prb,twocolumn,showpacs]{revtex4}
\usepackage{graphicx}
\begin{document}
\title{PES and XAS Study on
	Electronic Structures of Multiferroic RMnO$_3$ (R=Y, Er)}
\author {J.-S. Kang$^{*,1,2}$, S. W. Han$^{2}$, J.-G. Park$^{2,3}$, 
	S. C. Wi$^{1}$, S. S. Lee$^{1}$, G. Kim$^{1}$, 
	H. J. Song$^{4}$, H. J. Shin$^{4}$, W. Jo$^{5}$, 
	and B. I. Min$^{6}$} 
\affiliation{
$^{1}$Department of Physics, The Catholic University of Korea,
        Puchon 420-743, Korea\\
$^{2}$CSCMR, Seoul National University, Seoul 151-742, Korea\\
$^{3}$Department of Physics, Sungkyunkwan University,
        Suwon 440-746,, Korea\\
$^{4}$Pohang Accelerator Laboratory (PAL), 
	POSTECH, Pohang 790-784, Korea\\
$^{5}$Department of Physics, Ewha Womans University, 
	Seoul 120-750, Korea \\
$^{6}$Department of Physics, POSTECH, Pohang 790-784, Korea}

\date{\today}
\begin{abstract}
Electronic structures of multiferroic 
RMnO$_3$ (R=Y, Er) have been investigated 
by employing photoemission spectroscopy (PES) and x-ray absorption
spectroscopy (XAS).
We have found that Mn ions in RMnO$_3$ are 
in the trivalent high-spin state with the total spin of $S=2$.
The occupied Mn ($d_{xz} - d_{yz}$) states lie 
deep below $\rm E_F$, while the occupied 
Mn ($d_{xy} - d_{x^2 -y^2}$) states overlap very much 
with the O $2p$ states. 
It is observed that the PES spectral intensity of Mn $3d$ states 
is negligible above the occupied O $2p$ bands, suggesting that 
YMnO$_3$ is likely to be a charge-transfer insulator.
The Mn $d_{3z^2 -r^2}$ state is mostly unoccupied 
in the ferroelectric phase of YMnO$_3$.

\end{abstract} 
\maketitle

Hexagonal yttrium (Y) and rare-earth (R) manganites of the formula
RMnO$_3$ (R=Ho, Er, Tm, Yb, Lu, or Y) belong to an interesting class,
known as multiferroic materials, in which the ferroelectric and
magnetic ordering coexist at low temperatures
\cite{Bertaut63,Yakei63,Smolenskii82,Schmid94}.   
Hexagonal RMnO$_3$ compounds have antiferromagnetic ordering
with the Neel temperature (T$_N$) of  T$_N < 70-130$ K 
\cite{Bertaut63,Koehler64},   
and the ferroelectric ordering occurs at a high temperature 
(T$_E \sim 600-990$ K) \cite{Bertaut63,Yakei63}.   
RMnO$_3$ crystallizes in two structural phases;
the hexagonal phase when the ionic radius of R is small,
and the orthorhombic phase when the ionic radius of R is rather large.
Note that the ferroelectric ordering occurs only in the hexagonal 
phase of RMnO$_3$, while the magnetic ordering occurs in both hexagonal 
and orthorhombic phases.
In the hexagonal structure, each Mn ion is surrounded 	
by three in-plane and two apical oxygen ions,
and so it is subject to a trigonal crystal field \cite{Kats01}.
These MnO$_5$ blocks are connected two-dimensionally through 
their corners, and the triangular lattice of Mn$^{3+}$ ions is formed. 
These hexagonal RMnO$_3$ compounds experience characteristic 
distortions such as tilting of MnO$_5$ blocks and the displacement of 
R$^{3+}$ ions along the $c$ axis, causing a ferroelectric polarization 
\cite{Kats01,Aken04}.

There seems to be a strong coupling between the ferroelectric 
and magnetic ordering in hexagonal RMnO$_3$ compounds.
For example, a critical change of dielectric constants at T$_N$ has
been reported for RMnO$_3$ polycrystalline samples 
\cite{Huang97,Iwata98,Kimura02}, suggesting a coupling  
between the ferroelectric and magnetic ordering.
Further, the coupled antiferromagnetic and ferroelectric domains 
have been observed in YMnO$_3$ \cite{Fiebig02}, and
the magnetic phase of HoMnO$_3$ has been observed to
be controlled by the electric field \cite{Fiebig04}.
In the optical study of LuMnO$_3$ \cite{Souch03},
a strong coupling of antiferromagnetism to the optical absorption
spectra has been observed.

Understanding the origin of the coexistence of magnetism and 
ferroelectricity in hexagonal RMnO$_3$ is a fundamental physics 
question, but has not been well understood yet.
The $d^0$-ness rule is generally accepted in ferroelectricity.
That is, a ferroelectric displacement of B cation in ABO$_3$ is
inhibited if the formal charge of B ion does not correspond to a
$d^0$ electron configuration due to the strong on-site Coulomb
interaction between $d$ electrons.
In contrast, the occupancy of transition-metal $d$ electrons
is crucial in the magnetic ordering.
Thus the simultaneous magnetic and ferroelectric ordering
in RMnO$_3$ seems to break the $d^0$-ness rule.
It was suggested that this multiferroic property is a result of 
the effective 1-dimensional (1-D) $d^0$-ness along the $c$-axis
\cite{Hill02,Spaldin03}.
According to the electronic structure calculations 
\cite{Medvedeva00,Qian01},
the Mn$^{3+}$ ($3d^4$) ion in YMnO$_3$ is not a Jahn-Teller ion
since the highest occupied $3d$ level ($d_{xy}-d_{x^2 -y^2}$)
is non-degenerate because of the trigonal symmetry of the surrounding 
oxygen ions.
The $d_{3z^2 -r^2}$ state is mostly unoccupied so that this 
effective 1-D $d^0$ orbital along the $c$-axis allows ferroelectricity 
to occur via the usual ligand-field stabilization mechanism.
This is a plausible idea, but has not been confirmed
experimentally yet.

In order to understand the origin of the multiferroicity
in hexagonal RMnO$_3$, it is important 
to investigate the electronic structures of hexagonal RMnO$_3$,
including the valence states of Mn ions, the character
of the lowest unoccupied $3d$ states of the Mn ion 
below and above the ferroelectric transition.
In principle, these investigations are possible by employing 
the polarization-dependent soft x-ray absorption spectroscopy (XAS)
and valence-band photoemission spectroscopy (PES) measurements.
PES and XAS are powerful experimental methods for providing direct 
information on the electronic structures of solids. 
In practice, however, these spectroscopic measurements are not easy
for these multiferroic materials. 
Normally these systems are good insulators having wide band gaps,
and so they are not good for electron spectroscopy studies.
For the polarization-dependent experiments, single crystals
are prerequisite.
Further, it is difficult to study the changes in the electronic 
structure across T$_E$ using these spectroscopy experiments because 
it is practically not compatible
with the ultra high vacuum required for the experiments
to achieve these high temperatures (T$_E \sim 600-990$ K).

In this paper, we report the valence-band PES, O $1s$ XAS,
and Mn $2p$ XAS study of polycrystalline RMnO$_3$ (R=Y, Er) samples
at room temperature which belongs to the paramagnetic 
ferroelectric phase. 
To our knowledge, this is the first reliable PES and XAS study 
on multiferroic samples \cite{Yi00}.
This study provides the information on the electronic structures 
of RMnO$_3$ (R=Y, Er) in their ferroelectric and paramagnetic phases 
even though we did not perform the polarization-dependent 
spectroscopy measurement across T$_E$ or T$_N$. 


Polycrystalline RMnO$_3$ samples (R=Y, Er) were synthesized
by using the standard solid-state reaction method.
Cation oxides of R$_2$O$_3$ (R=Y, Er) ($99.999 \%$) and 
Mn$_2$O$_3$ ($99.999 \%$) were thoroughly mixed in order to
achieve a homogeneous mixture.
The mixed powders were heated to $900 ^{\circ}$ C for 12 hours 
and later they were annealed at $1100 ^{\circ}$  for 24 hours 
and subsequently at $1200 ^{\circ}$ C for 24 hours 
before final sintering at $1350 ^{\circ}$ C for 24 hours 
with intermediate grindings.
The purpose of intermediate grindings was to prevent
the formation of impurity phases.
The x-ray diffraction (XRD) measurements at room temperature showed 
that all the samples have a single hexagonal RMnO$_3$ phase.

Valence-band PES, O $1s$ XAS, and Mn $2p$ XAS
measurements were performed at the 8A1 undulator beamline of the Pohang
Accelerator Laboratory (PAL).   
Samples were cleaned {\it in situ} by repeated scraping with a diamond 
file and the data were obtained at room temperature with the pressure 
better than $4 \times 10^{-10}$ Torr.
The Fermi level $\rm E_F$ \cite{calib} and the overall instrumental 
resolution (FWHM) of the system were determined from the valence-band 
spectrum of a scraped Pd metal in electrical contact with a sample.
The FWHM was about $100 - 400$ meV between a photon energy 
$h\nu\sim 130$ eV and $h\nu\approx 600$ eV.
All the spectra were normalized to the incident photon flux.     
The XAS spectra were obtained by employing the total 
electron yield method.
The experimental energy resolution for the XAS data was set
to $\sim 100$ meV at the O $1s$ and Mn $2p$ absorption thresholds
($h\nu \approx$ 500-600 eV).


\begin{figure}[t]
\includegraphics[scale = 0.45]{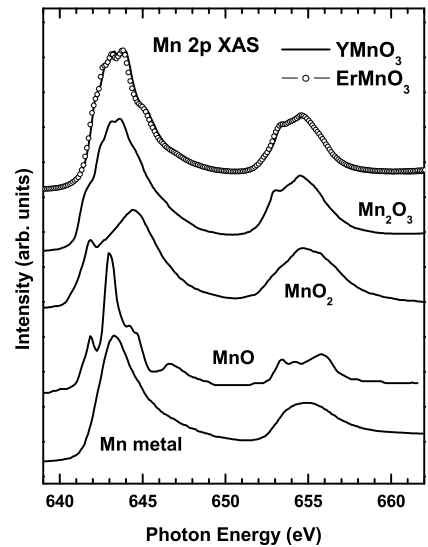}
\caption{Comparison of the Mn $2p$ XAS spectra of 
	YMnO$_3$ (solid lines) and ErMnO$_3$ (symbols) to those of 
	Mn$_2$O$_3$ (Mn$^{3+}$) [Ref.~\protect\cite{Ghigna01}], 
	MnO$_2$ (Mn$^{3+}$) [Ref.~\protect\cite{Mitra03}],
	MnO (Mn$^{2+}$) [Ref.~\protect\cite{Mitra03}], 
	and Mn metal [Ref.~\protect\cite{Yonamoto01}]. }
\label{2pxas} 
\end{figure}

Figure~\ref{2pxas} compares the Mn $2p$ XAS spectra
of RMnO$_3$ (R=Y, Er) to those of reference Mn compounds
having formal Mn valences of
$3+$ (Mn$_2$O$_3$, reproduced from Ref.~\cite{Ghigna01}), 
$4+$ (MnO$_2$ reproduced from Ref.~\cite{Mitra03}), 
$2+$ (MnO reproduced from Ref.~\cite{Mitra03}), 
and that of Mn metal (reproduced from Ref.~\cite{Yonamoto01}). 
The XAS data of MnO$_2$, MnO, and Mn metal were shifted by 
$-0.8$ eV, $+1.0$ eV, and $-1.3$ eV, respectively,
to allow for the better comparison of the data. 
It is well known that the peak positions and the line shape 
of the Mn $2p$ XAS spectrum depend on the local electronic structure 
of the Mn ion, so that the $2p$ XAS spectrum provides the information 
on the valence state of the Mn ion \cite{Groot90,Laan92}.
Figure~\ref{2pxas} shows clearly that the Mn $2p$ XAS spectra 
of RMnO$_3$ (R=Y, Er) are essentially identical to each other, 
and that they are very similar to that of Mn$_2$O$_3$.
For comparison,	
they are quite different from those of MnO ($2+$), MnO$_2$ ($4+$), 
and Mn metal. This observation indicates that the valence states of 
Mn ions in RMnO$_3$ are nearly trivalent (Mn$^{3+}$), with the $3d^4$ 
configuration, but far from being divalent (Mn$^{2+}$, $3d^5$) or 
tetravalent (Mn$^{4+}$, $3d^3$). 
This finding is consistent with the finding of Mn K-edge XANES
\cite{Wu03}, and  with the general consensus of Mn$^{3+}$ ions 
in the ionic bonding picture for hexagonal RMnO$_3$. 

\begin{figure}[t]
\includegraphics[scale = 0.45]{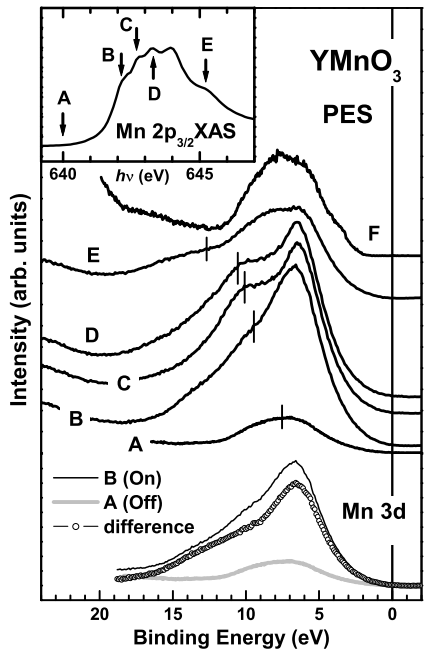}
\caption{Valence-band PES spectra of YMnO$_3$ 
        near the Mn $2p_{3/2} \rightarrow 3d$ absorption edge. 
	Inset: The Mn $2p_{3/2}$ XAS spectrum of YMnO$_3$.
	Arrows denote $h\nu$'s where the valence-band PES spectra 
        were obtained.
        Bottom: Comparison of the on-resonance (solid line) 
        and off-resonance valence-band PES spectra (gray line) 
        in Mn $2p \rightarrow 3d$ RPES, and the difference 
	between these two (open dots). }
\label{rpes}
\end{figure}

\begin{figure}[t]
\includegraphics[scale = 0.4]{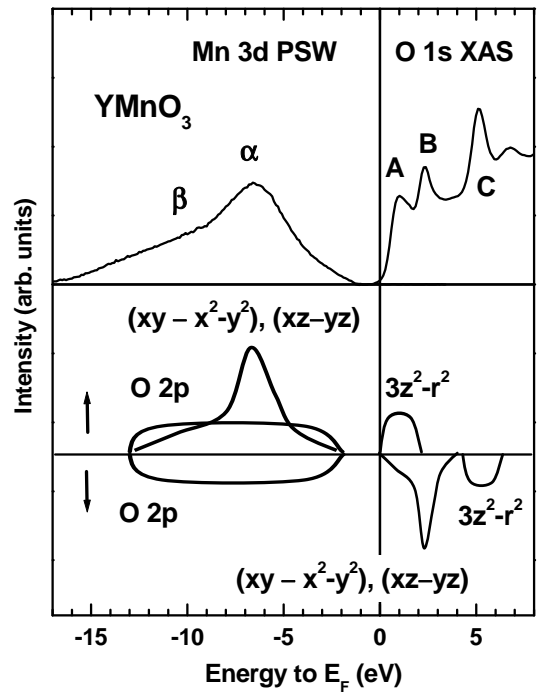}
\caption{Top: Combined Mn $3d$ PSW and O $1s$ XAS for YMnO$_3$. 
        Bottom: The schematic diagram for the Mn $3d$ PDOS 
	of YMnO$_3$.  $\uparrow$ and $\downarrow$ represent 
	the majority- and minority-spin states, respectively.  }
\label{pdos}
\end{figure}

If Mn ions in RMnO$_3$ have the $3d^4$ configurations, 
then the next question would be which states are occupied
which ones are unoccupied, and where these states are located 
with respect to $\rm E_F$.
In order to answer these questions, we have investigated
the photoemission spectral weight distribution of the Mn $3d$
electrons, by employing resonant photoemission spectroscopy (RPES) 
near the Mn $2p \rightarrow 3d$  absorption threshold
\cite{Kang02}.

Top of Fig.~\ref{rpes} shows the valence-band RPES spectra of 
YMnO$_3$ near the Mn $2p_{3/2}$ absorption edge. 
We have also done RPES measurements for ErMnO$_3$, but the 
valence-band PES spectra have large contribution from
Er $4f$ electron emissions, which overlap with both Mn $3d$ and
O $2p$ emissions. So we do not present the PES data for ErMnO$_3$
in this paper. The inset shows the Mn $2p_{3/2}$ XAS spectrum 
of YMnO$_3$, and the arrows in the XAS spectrum represent 
$h\nu$'s where the valence-band spectra were obtained.
The off-resonance valence-band PES spectrum ($A$) consists
of both O $2p$ and Mn $3d$ electron emissions with comparable
contributions, but has the negligible contribution from Y $s/p$ 
electron emissions \cite{Yeh85}.
The valence-band PES spectrum, labeled as $F$, was obtained with 
$h\nu=1486.6$ eV. This spectrum is very similar to 
the off-resonance spectrum ($A$), because, at $h\nu=1486.6 $eV, 
the O $2p$ and Mn $3d$ electron emissions 
are also of comparable magnitudes \cite{Yeh85}.
The enhanced features near $\sim 7$ eV binding energy
at the Mn $2p \rightarrow 3d$ absorption energy ($B$)
represent the resonant Mn $3d$ electron emission. 
Therefore the difference between the on-resonance and off-resonance 
spectra can be considered to represent the {\it bulk} Mn $3d$ partial 
spectral weight (PSW) distribution \cite{Kang02}.
The vertical bars, marked for those features
that shift away from $\rm E_F$ with increasing $h\nu$, 
denote the Mn LMM Auger emission that appears at a fixed 
kinetic energy (KE) of KE $\sim 635$ eV \cite{AugerHB95}.
These Auger peaks also reveal the intensity enhancement near 
the Mn $2p$ absorption threshold.
It is known that the transition-metal (T) Auger peaks show 
the resonant behavior
near the T $2p \rightarrow 3d$ RPES \cite{Kang03}.

Bottom of Fig.~\ref{rpes} presents the extraction procedure of 
the Mn $3d$ PSW for YMnO$_3$. 
As a first approximation, it is taken as the difference  between the 
Mn $2p \rightarrow 3d$ on-resonance spectrum (solid line) and 
off-resonance spectrum (gray line).  In this extraction procedure, 
we have used the on-resonance spectrum at B,
instead of C or D, to reduce the effect of the Mn LMM Auger emission.
The extracted Mn $3d$ PSW exhibits a peak centered at 
$\sim 7$ eV binding energy with an asymmetric and long tail 
to the high binding energy side (up to $\sim 15$ eV).
Part of this high binding tail is ascribed to 
the underlying Mn Auger peak. 
The extracted Mn $3d$ PSW for YMnO$_3$ shows that Mn $3d$ states 
are located well below $\rm E_F$ but that there is nearly
no Mn $3d$ occupied states near $\rm E_F$.
Note that both the off-resonance spectrum (A) and
the $h\nu=1486.6$ eV spectrum (F)
represent the mixture of O $2p$ and Mn $3d$ emissions. 
Therefore the bottom part of Fig.~\ref{rpes}
reveals that the occupied Mn $3d$ states lie in the middle of 
the O $2p$ states and that they overlap with the O $2p$ states 
very much, implying the strong hybridization between 
Mn $3d$ and O $2p$ states.

Figure~\ref{pdos} shows the combined Mn $3d$ PSW 
and the O $1s$ XAS spectrum of YMnO$_3$ \cite{shift}. 
The O $1s$ XAS spectrum represents the transition from the O $1s$ 
core level to the unoccupied O $2p$ states which are hybridized 
with the other electronic states, so it provides 
a reasonable estimate of the unoccupied conduction bands.
The peaks $A$, $B$, $C$ in the O $1s$ XAS spectrum of YMnO$_3$  
have also been observed in ErMnO$_3$ and they were very similar
to each other, indicating that they have mainly
the O $2p-$Mn $3d$ character.
At the bottom of Fig.~\ref{pdos}, we also provide a schematic diagram 
for the partial densities of states (PDOS) for YMnO$_3$.
We have determined this schematic PDOS diagram 
through the comparison 
of the PES/XAS data and the LSDA+$U$  
(LSDA: local spin-density approximation, $U$: Coulomb interaction)
band calculations for YMnO$_3$ \cite{Medvedeva00,Qian01}.

The features observed in the PES/XAS are ascribed to 
the following states.
$\alpha$ : the occupied Mn $3d$ states consisting of 
the narrow ($d_{xz}\uparrow - d_{yz}\uparrow$) states,
which are superposed on top of the rather broad 
($d_{xy}\uparrow - d_{x^2 -y^2}\uparrow$) states,
$\beta$ (the high binding energy shoulder to $\alpha$):
the Mn Auger peak,
$\alpha+\beta$ (the broad valence-band PES features underneath
the Mn $3d$ states): the O $2p$ states.
Note that the occupied O $2p$ states are hybridized with 
($d_{xy} - d_{x^2 -y^2}$) states, consistent with
the band-structure calculations \cite{Medvedeva00,Qian01}.
In the XAS side, the marked peaks represent the following states.
$A$: the unoccupied Mn $d_{3z^2 -r^2}\uparrow$ states,
$B$: the unoccupied Mn ($d_{xy}\downarrow - d_{x^2 -y^2}\downarrow$)
and ($d_{xz}\downarrow - d_{yz}\downarrow$) states,
$C$: the unoccupied Mn $d_{3z^2 -r^2}\downarrow$ states.
Thus the state just above $\rm E_F$ corresponds to Mn 
$d_{3z^2 -r^2}\uparrow$,
and so the trivalent  Mn$^{3+}$ ions in RMnO$_3$ are 
in the high-spin state ($S=2$) with the configuration of
$(d_{xz}\uparrow -d_{yz}\uparrow)^2$$(d_{xy}\uparrow 
- d_{x^2 -y^2}\uparrow)^2$.
Considering the uncertainty in the binding energy calibration of 
PES data \cite{calib}, the energy separation between the peak 
$\alpha$ in the valence-band PES and the peak A in the O $1s$ 
XAS amounts to $7 \sim 8$ eV, which gives a rough measure of 
$U + \Delta_{CF}$ ($U$: on-site Coulomb interaction  
between Mn $3d$ electrons, $\Delta_{CF}$: the crystal-field splitting
between $d_{3z^2 -r^2}$ and $d_{xz} - d_{yz}$).
Using $\Delta_{CF}\approx 2$ eV from the LSDA band structure 
calculation \cite{Medvedeva00},
the estimated value of $U$ would be of the order of $5 \sim 6$ eV. 
The energy separation ($\sim 2$ eV) between the top of the valence 
band and the peak $A$ in the O $1s$ XAS is expected to correspond 
to the lowest energy peak in the optical absorption spectrum 
for YMnO$_3$ \cite{Yi00,Kala03}.
We ascribe this energy separation to the energy difference between 
the occupied O $2p$ states, which are strongly hybridized 
to the ($d_{xy}\uparrow - d_{x^2 -y^2}\uparrow$) states,
and the unoccupied Mn $d_{3z^2 -r^2}\uparrow$ states.

Note that the extracted Mn $3d$ PSW shows negligible 
spectral weight near $\rm E_F$.
Thus our schematic PDOS diagram shown in Fig.~\ref{pdos} implies
that the topmost electronic states in the valence bands of YMnO$_3$ 
(those closest to $\rm E_F$) are mostly O $2p$ states.
According to this picture, the lowest-energy optical transition 
would be the O $p$-Mn $d$ transition, suggesting that YMnO$_3$ 
is likely to be a charge-transfer insulator.  
This picture is consistent with the  electronic structures obtained
by using the LSDA+$U$, rather than those by the LSDA method 
\cite{Medvedeva00}.
The present data, however, indicate that the Mn $3d$ Coulomb interaction 
$U$ is not as large as employed in Ref. \cite{Medvedeva00}, $U=8$ eV.
Noteworthy is that our model for the electronic structure of YMnO$_3$
is different from that proposed based on the optical
study of LuMnO$_3$ \cite{Souch03}.
They observed a sharp peak at 1.7 eV in the optical absorption
spectrum for LuMnO$_3$ even at room temperature, which was ascribed 
to the on-site Mn $d$-$d$ transition.
In contrast, the similar absorption peaks were observed in RMnO$_3$
(R=Sc, Y, Er), which were interpreted as arising from charge transfer 
from O $2p$ to Mn $3d$ states \cite{Kala03}.
The present PES/XAS data indicate that the latter interpretation
is more consistent with the real electronic structures in
RMnO$_3$.


In conclusion, the electronic structures of 
hexagonal multiferroic RMnO$_3$ (R=Y, Er) materials
have been investigated by employing Mn $2p \rightarrow 3d$ RPES,
Mn $2p$ XAS, and O $1s$ XAS. 
The Mn $2p$ XAS spectra of RMnO$_3$ (R=Y, Er) show that  
Mn ions are in the formally trivalent Mn$^{3+}$ states, 
implying the $(d_{xz}\uparrow -d_{yz}\uparrow)^2$$(d_{xy}\uparrow 
- d_{x^2 -y^2}\uparrow)^2$ configurations 
with the total spin of $S=2$ per Mn ion. 
According to Mn $2p \rightarrow 3d$ RPES for YMnO$_3$, 
the occupied Mn ($d_{xz} - d_{yz}$) states lie very deep 
below $\rm E_F$, with a peak around $\sim 7$ eV binding energy.
The occupied Mn ($d_{xy} - d_{x^2 -y^2}$) states overlap with 
the O $2p$ states very much, but show negligible Mn $3d$ spectral 
weight above the O $2p$ states, suggesting that YMnO$_3$ is 
likely to be a charge-transfer insulator.
The lowest unoccupied peak in the O $1s$ XAS is ascribed 
to the unoccupied Mn $d_{3z^2 -r^2}$ state.
This finding is compatible with the recent structural studies of
RMnO$_3$ \cite{Kats01,Aken04}, which show tilting of MnO$_5$ blocks 
in their ferroelectric phases.

Acknowledgments$-$
We thank S.-W. Cheong for helpful discussions. 
This work was supported by the KRF (KRF-2002-070-C00038) and 
by the KOSEF through the CSCMR at SNU and the eSSC at POSTECH.
The PAL is supported by the MOST and POSCO in Korea.


\begin{thebibliography}{99}
\bibitem{calib} Since RMnO$_3$ samples are insulating,
        there is some ambiguity in determining $\rm E_F$.
  	However, no time-dependent charging effects were observed 
	in the valence-band PES spectra of RMnO$_3$. 
	Therefore the absolute values of binding energies
	do not have much meaning in these data,
	but the relative energy separations are meaningful.
\bibitem{shift} In this comparison, the inelastic background 
	has been subtracted in the Mn $3d$ PSW, and
	the O $1s$ XAS spectrum has been shifted by $-530$ eV 
	by referring to the optical gap size of YMnO$_3$.
\end{thebibliography}
\end{document}